\documentclass[9pt]{article}
\usepackage{spconf,algorithm,algpseudocode,subfigure,amsfonts, amsmath,amssymb,graphicx,cite,color,nicefrac,siunitx}

\allowdisplaybreaks

\def\tH{\pmb{\mathcal{H}}}

\def\bbe{\boldsymbol{\beta}}

\def\({\left(}
\def\){\right)}
\def\[{\left[\,}
\def\]{\,\right]}

\def\0{\boldsymbol{0}}
\def\1{\boldsymbol{1}}

\def\A{\mathbf{A}}

\def\b{\mathbf{b}}
\def\B{\mathbf{B}}

\def\bC{\mathbb{C}}

\def\D{\mathbf{D}}

\def\G{\mathbf{G}}

\def\H{\mathbf{H}}

\def\I{\mathbf{I}}

\def\n{\mathbf{n}}

\def\s{\mathbf{s}}

\def\v{\mathbf{v}}
\def\V{\mathbf{V}}

\def\x{\mathbf{x}}

\def\diag{\mathrm{diag}}

\newtheorem{theorem}{Theorem}

\title{Tensor-Based Parameter Estimation of Double Directional Massive MIMO Channel with Dual-Polarized Antennas}

\name{Cheng Qian$^{\dagger}$ \qquad Xiao Fu$^\star$ \qquad Nicholas D. Sidiropoulos$^\dagger$ \qquad Ye Yang$^{\ddagger}$
\thanks{Author e-mails: alextoqc@gmail.com, xiao.fu@oregonstate.edu, nikos@virginia.edu, yangye@huawei.com.}
}

\address{
$^{\dagger}$Dept. of Electrical and Computer Engineering, University of Virginia, Charlottesville, VA 22904\\
$^{\star}$School of Electrical Engineering and Computer Science, Oregon State University, Corvallis, OR 97331\\
$^\ddagger$ Physical Layer \& RRM IC Algorithm Dept., WN Huawei Co., Ltd, Shanghai, China}

\begin{document}

\maketitle

\begin{abstract}
    The 3GPP suggests to combine dual polarized (DP) antenna arrays with the double directional (DD) channel model for downlink channel estimation. This combination strikes a good balance between high-capacity communications and parsimonious channel modeling, and also brings limited feedback schemes for downlink channel estimation within reach. However, most existing channel estimation work under the DD model has not considered DP arrays, perhaps because of the complex array manifold and the resulting difficulty in algorithm design. In this paper, we first reveal that the DD channel with DP arrays at the transmitter and receiver can be naturally modeled as a low-rank four-way tensor, and thus the parameters can be effectively estimated via tensor decomposition algorithms. To reduce computational complexity, we show that the problem can be recast as a four-snapshot three-dimensional harmonic retrieval problem, which can be solved using computationally efficient subspace methods. On the theory side, we show that the DD channel with DP arrays is identifiable under very mild conditions, leveraging identifiability of low-rank tensors. Numerical simulations are employed to showcase the effectiveness of our methods.
\end{abstract}


\section{Introduction}

The dual-polarized (DP) antenna array has many appealing features and is thus considered a key technique for next generation communications and massive MIMO \cite{3gpp,3gpp_stand,polarizationHeath}.
For example, Foschini and Gans \cite{polarization1} showed that the capacity for systems with DP antennas at the transmitter can be increased up to 50\% compared to systems without polarization. 
Besides the increased capacity, DP antennas have other key advantages such as small size, easy installation, good interference mitigation performance, high link reliability, and high ability of interference filtering, just to name a few \cite{3gpp_stand,polarization1,3gpp,polarizationHeath}.

In the recent releases of technical specifications suggested by 3GPP, the DP array and the double directional (DD) channel model are considered key techniques \cite{3gpp}. The DD channel model is parsimonious for multipath channels with a small number of dominant paths, and parsimony is really essential for designing limited feedback schemes for downlink channel estimation in massive MIMO \cite{3gpp,3gpp_stand,polarizationHeath,panos}. Specifically, 3GPP suggests that the mobile users estimate the DD channel parameters such as directions-of-arrival (DOAs), directions-of-departure (DODs), the path loss associated with each path and the polarization angles of the DP array, and then feed back these parameters to the base station (BS). This strategy is rather economical, as it is expected that the number of dominant paths will be small to moderate in practical deployments.
On the other hand, to the best of our knowledge, there is very limited work related to the DD-DP parameter estimation problem. Most of the existing channel estimation algorithms such as \cite{jarvis,codebook4,panos,ce1} do not take polarization into consideration, and thus cannot be applied to this particular system. The early algorithm proposed in \cite{polarized1} in the context of array processing can only handle a small number of paths, since the maximum number of identifiable paths in \cite{polarized1} is restricted by the size of the receive array. 
\vspace{-.1em}

In this work, we focus on the parameter estimation problem under the DD channel model with DP arrays. 
Specifically, we first show that the DD channel with DP arrays at the transmitter and receiver can be naturally modeled as a low-rank four-way tensor. Leveraging this structure, we recast the associated parameter estimation problem as a Parallel Factor Analysis (PARAFAC) decomposition problem \cite{Sid2017} and handle it using effective tensor decomposition algorithms.
To reduce computational complexity, we also formulate channel estimation as a three-dimensional (3-D) harmonic retrieval problem, which can be solved by a computationally efficient subspace method, namely, the \emph{improved multidimensional folding (IMDF)} method \cite{liu}. On the theory side, we show that the channel and polarization parameters are identifiable under very mild and practical conditions -- even when the number of paths largely exceeds the number of receive antennas, a practically important case that classic DP channel estimation algorithms as in \cite{polarized1} cannot cope with.
Simulations are provided to showcase the effectiveness of the proposed methods.
%

\section{Signal Model}
We consider a massive MIMO system, where there is one BS equipped with an $M_x\times M_y$ DP uniform rectangular array (URA) and one mobile station (MS) with an $M_r$-element DP uniform linear array (ULA), which is a practical setting that is of interest to industry \cite{3gpp}. Throughout the paper, we consider DP array elements consisting of a pair of crossed dipoles. In the literature, this type of DP array is also known as a ``cross-polarized'' array \cite{polarized1}. The number of transmit antennas is $M_t=M_xM_y$. The signal received by the user is given by 
\begin{align}\label{eq:channel}
\x(t) = \H\s(t) + \n(t),\ t=1,\cdots,N
\end{align}
where $\s(t)\in\bC^{2M_t \times 1}$ is the transmitted signal, $\n(t)$ is zero-mean i.i.d.circularly symmetric complex Gaussian noise.
By properly arrange elements, the downlink channel matrix can be represented as the following four-block matrix:
\begin{align}\label{H}
\H = \begin{bmatrix} \H^\mathrm{(V_r,V_t)} & \H^\mathrm{(V_r,H_t)} \\ \H^\mathrm{(H_r,V_t)} & \H^\mathrm{(H_r,H_t)}  \end{bmatrix} \in\bC^{2M_r\times 2M_t}.
\end{align}
where $\H^\mathrm{(V_r,V_t)}\in\bC^{M_r\times M_t}$ is a channel matrix between all the V-polarized transmit antennas and V-polarized receive antennas, and $\H^\mathrm{(V_r,H_t)}\in\bC^{M_r\times M_t}$ is a channel matrix between all the H-polarized transmit antennas and V-polarized receive antennas; likewise for the other two blocks in \eqref{H}.

For notational simplicity, let $\mathrm{p}\in\{\mathrm{V_r,H_r}\}$ and $\mathrm{q}\in\{\mathrm{V_t,H_t}\}$. Then, according to \cite{3gpp_stand}, the $(\mathrm{p,q})$ subchannel matrix is modeled as 
\begin{align}
\H^\mathrm{(p,q)} = \V_r\diag\big(\bbe^\mathrm{(p,q)}\big)\V_t^H
\end{align}
where $(\cdot)^H$ is the conjugate transpose,
$\V_r = [ \v_{r}(\theta_{1}) \cdots \v_{r}(\theta_{K}) ]$, 
$\V_t=[ \v_{t}(\vartheta_{1},\varphi_{1}) \cdots \v_{t}(\vartheta_{K},\varphi_{K}) ]$, and 
$\bbe^\mathrm{(p,q)} =
[\beta_1^\mathrm{(p,q)}  \cdots  \beta_K^\mathrm{(p,q)} ]^T$ stands for path-losses with $(\cdot)^T$ denoting the transpose. Note that $\{\theta_k\}$ are DOAs, $\{\vartheta_k\}$ and $\{\varphi_k\}$ are azimuth and elevation DODs, respectively. Throughout of this paper, we assume that the receive and transmit antennas have the same half-wavelength inter-element spacing. Then we have $[\v_{r}(\theta_k)]_m=e^{j\pi(m-1)\sin(\theta_k)}$ and $\v_t(\vartheta_k,\varphi_k) = \v_{y,k}\otimes\v_{x,k}$, where $[\x]_i$ denotes the $i$th element of $\x$, $[\v_{x,k}]_{l_x} = e^{j\pi (l_x-1)\sin(\varphi_k)\cos(\vartheta_k)}, l_x = 0,\cdots,M_x-1$ and $[\v_{y,k}]_{l_y} = e^{j\pi (l_y-1)\sin(\varphi_k)\sin(\vartheta_k)},l_y=0,\cdots,M_y-1$.

Now the channel matrix in \eqref{H} can be rewritten as
\begin{align}\label{eq:3-Dmimo}
\!\!\!\!\H = 
\begin{bmatrix}
\V_r\diag\big(\bbe^\mathrm{(V_r,V_t)}\big)\V_t^H \!&\!\! \V_r\diag\big(\bbe^\mathrm{(V_r,H_t)}\big)\V_t^H \\
\V_r\diag\big(\bbe^\mathrm{(H_r,V_t)}\big)\V_t^H \!&\!\! \V_r\diag\big(\bbe^\mathrm{(V_t,H_t)}\big) \V_t^H
\end{bmatrix}.
\end{align}
In this model, to determine the channel ${\bf H}$, we only need to estimate $K$ DOAs, $K$ azimuth angles, $K$ elevation angles and $4K$ complex path-losses.  Compared to the size of the channel, which is $4M_rM_t$, such parameterization is rather economical and is suitable for massive MIMO downlink channel estimation and limited feedback where both $M_t$ and $M_r$ (especially $M_t$) can be very large.



\subsection{Challenges}
Although we have explicitly written down the channel model in \eqref{eq:channel}, how to effectively estimate the parameters of interest is still unclear.
Specifically, assume that ${\bf H}$ can be estimated at the receiver by matched filtering, i.e., ${\bf H}={\bf X}{\bf S}^H$ under a pre-selected row-orthogonal pilot sequence ${\bf S}$, where ${\bf X}=[{\bf x}(1),\ldots,{\bf x}(N)]$ and ${\bf S}=[{\bf s}(1),\ldots,{\bf s}(N)]$. Estimating the DOA, DOD and path-loss parameters is still very challenging.
One popular type of technique to estimate parameters of the (non-DP) DD channel is described in \cite{jarvis,codebook4,panos}, where the DOA and DOD domains are descretized to fine angle grids using two overcomplete angle dictionaries (codebooks), denoted by ${\bf D}_t$ and ${\bf D}_r$. Then, we have $\H\approx (\I_2\otimes\D_r)\G(\I_2\otimes\D_t)^H$, where ${\bf G}$ is a sparse matrix that selects out the columns associated with the active DODs and DOAs from the dictionaries. This way, the parameter estimation problem becomes a sparse recovery problem that can be handled by formulations such as $\min_{\bf g}~\|{\bf h}-(\I_4\otimes{\bf D}_t^\ast \otimes {\bf D}_r) {\bf g}\|_2^2 + \lambda \|{\bf g}\|_1$,
where ${\bf h}={\rm vec}({\bf H})$ with $\text{vec}(\cdot)$ being the vectorization operator and ${\bf g}={\rm vec}({\bf G})$; and other sparse optimization algorithms such as orthogonal matching pursuit . 

The difficulty is that to ensure good spatial resolution, both ${\bf D}_t\in\mathbb{C}^{M_t\times D_t}$ and ${\bf D}_r\in\mathbb{C}^{M_r\times D_r}$ are very ``fat'' matrices, where $D_t$ and $D_r$ denotes the number of angle grids after quantization. Consequently, $(\I_4\otimes{\bf D}_t^\ast \otimes {\bf D}_r)$ is $4M_tM_r\times 4D_tD_r$. If one quantizes the DOA and DOD space (ranging from $-90^\circ$ to $90^\circ$) using a resolution of one degree, then $4D_tD_r=131,044$ -- which poses an extremely hard sparse optimization problem. Many compromises, such as coarse quantization and hierarchical or hybrid algorithms \cite{codebook4,panos}, have been employed to circumvent this issue in the literature. However, when two-dimensional antennas are deployed in both receiver and transmitter, the dictionary size can reach $4(D_tD_r)^2$, which is hopeless.  

\section{Proposed Approach}

\subsection{Tensor-Based Method and Identifiability}
Our proposed approach starts by noticing that ${\bf H}$ is in fact a four-way tensor of rank (at most) $K$; to see this, vectorize the four blocks in ${\bf H}$ and then stack them in a tall matrix, such that we have
\begin{align}\label{h}
    \breve{\H}= \(\V_{y}^*\odot\V_x^*\odot\V_r\)\B^T
\end{align}
where $(\cdot)^*$ denotes conjugation, $\odot$ is the Khatri-Rao product, $\V_x = [\v_{x,1} \cdots \v_{x,K}]$,
$\V_y = [\v_{y,1} \ \cdots\ \v_{y,K}]$ and
$\B = [\bbe^\mathrm{(V_r,V_t)}\; \bbe^\mathrm{(V_r,H_t)}\; \bbe^\mathrm{(H_r,V_t)}\; \bbe^\mathrm{(H_r,H_t)}]^T\in\bC^{4\times K}$.
Note that \eqref{h} is the definition of a four-way tensor of rank $\leq K$ in matrix form \cite{Sid2017}.


By noticing the tensor structure of ${\bf H}$, various tensor decomposition algorithms such as those in \cite{parafac1,parafac2} can be directly applied to estimate ${\bf V}_x$, ${\bf V}_y$, ${\bf V}_r$ and $\B$ via solving the following:
\begin{align}\label{Prob:H1}
\min_{\V_r,\V_x,\V_y,\B} \left\| \check{\H} -  \left(\V_y^*\odot\V_x^*\odot\V_r\right)\B^T \right\|_F^2
\end{align}
where $\|\cdot\|_F$ is the Frobenius norm. 
Note that a salient feature of tensors is that the factors are uniquely identifiable under mild conditions, as we will explain shortly. Once ${\bf V}_x$, ${\bf V}_y$, ${\bf V}_r$ and $\B$ are estimated, the parameters $\{\hat\theta_k,\hat{\vartheta}_k, \hat{\varphi}_k, \bbe^{(\rm p,q)}\}$ can be computed in closed-form.
Since $\v_{x,k},\v_{y,k}$ and $\v_{r,k}$ are Vandermonde vectors, we may use 
\begin{align}
\hat{\theta}_k &= \sin^{-1}\( \frac{1}{\pi } \angle(\overline{\hat\v}_{r,k}^H\underline{\hat\v}_{r,k})  \) \\
\hat{\varphi}_k &= \sin^{-1}\( \frac{1}{\pi} \sqrt{\big(\angle(\overline{\hat\v}_{x,k}^H\underline{\hat\v}_{x,k})\big)^2 \!\!+ \big(\angle(\overline{\hat\v}_{y,k}^H\underline{\hat\v}_{y,k})\big)^2}  \)  \\
\hat{\vartheta}_k &= \tan^{-1}\( \nicefrac{\angle(\overline{\hat\v}_{y,k}^H\underline{\hat\v}_{y,k}) }{\angle(\overline{\hat\v}_{x,k}^H\underline{\hat\v}_{x,k})} \)
\end{align}
where $\angle(\cdot)$ takes the phase of its argument, $\overline{\x}$ and $\underline{\x}$ are the vectors consisting of the first and last $(M-1)$ entries of $\x$ with length $M$, respectively. Any other single-tone frequency estimation algorithm, e.g., \cite{freqest1,freqest2} or ML-based (periodogram) methods can also be used, for better accuracy.

We should mention that by solving \eqref{Prob:H1} using any of the existing tensor decomposition algorithms, we already have an initial estimate of ${\B}$, i.e., the path-losses.
However, since there is an intrinsic scaling ambiguity of tensor decomposition, such an initial estimate may not be useful. Nevertheless, this issue is easy to fix. Note that the array manifolds $\hat{\A}_r,\hat{\A}_{x},\hat{\A}_y$ {\it without scaling ambiguity} can be constructed from $\{\hat{\theta}_k,\hat{\varphi}_k,\hat{\vartheta}_k\}_{k=1}^K$. 
Then, the estimate of ${\B}$ without scaling ambiguity can be computed from the following LS problem:
\begin{align}\label{eq:recal_B}
\hat{\B} &\leftarrow \arg\min_{\B}
\left\|\breve{\H}-(\hat\V_y^*\odot\hat\V_x^*\odot\hat\V_r) \B^T\right\|_F^2.
\end{align}

In terms of theoretical guarantees of identifiability, we have the following theorem:


\begin{theorem}
	The proposed approach can uniquely identify the parameters of interest under the DD channel model with DP arrays provided that $\min{(M_r,K)} + \min(M_x,K) + \min(M_y,K) + \min{(4,K)} \geq 2K + 3$.
\end{theorem}

One can easily check that $\{\V_{x},\V_{y},\V_r,\B\}$ meet the $k$-rank condition \cite{krank} provided that all the DOA, DOD and path-loss are not the same, which is a mild condition considering the random nature of multi-path. Thus, Theorem 1 essentially follows from \cite{nikos1}. Much better results can also be claimed, albeit in the almost surely sense -- see \cite{Sid2017}.

\subsection{IMDF and Identifiability}
The `naive' tensor-based method ignores the Vandermonde structure of some of the array manifold vectors in its first step, only to impose it later. This is suboptimal.  Theorem~1 in particular is a general bound that neglects the Vandermonde structure in $\v_x,\v_y$ and $\v_r$. If we take this structure into account, a better uniqueness condition can be obtained. 
To this end, we rearrange the elements of $\breve{\H}$ such that the resulting tensor is with dimension $M_r\times M_x\times M_y\times 4$, i.e., $\sum_{k=1}^{K} \v_{r,k}\circ\v_{x,k}^*\circ\v_{y,k}^*\circ\b_k$,
where $\circ$ denotes the outer product and $\b_k$ is the $k$th column of $\B$.
The above can be viewed as a multi-snapshot 3-D harmonic retrieval problem, where the number of snapshots is four, and each snapshot is written as 
\begin{align}\label{tildetH}
\tH^{(\rm p,q)} = \sum_{k=1}^{K} \beta_k^{(\rm p,q)} \v_{r,k}\circ\v_{x,k}^*\circ\v_{y,k}^*,~ \forall \mathrm{p,q}.
\end{align}


\begin{theorem}\label{theorem2}
The parameters $\big\{\theta_k,\varphi_k,\vartheta_k,\beta_k^{(\rm p,q)}\big\}$ are all uniquely identifiable by the IMDF based procedure provided that 
	\begin{align}
		K\leq \arg\max_{F,P_r,P_x,P_y} &~F\notag\\
		\mathrm{s. t.}\quad & \max\Big((P_r-1)P_xP_y, P_r(P_x-1)P_y, \notag\\
										&\quad\qquad P_rP_x(P_y-1)\Big) \geq F \notag\\
										& 8Q_rQ_xQ_y \geq F
	\end{align}
	where $P_r+Q_r=M_r+1, P_x+Q_x=M_x+1,P_y+Q_y=M_y+1$.
\end{theorem}

This follows by invoking the identifiability result for the IMDF algorithm for multi-dimensional harmonic retrieval \cite{liu}, which is far stronger compared to that in Theorem~1. For example, when $M_x=4,M_y=8,\text{and }M_r=2$, the identifiability of Theorem~1 is $K=7$, while the identifiability of Theorem 2 is $K=32$.
Furthermore, even when the MS only has a single dual-polarized antenna, it can be shown using the IMDF based approach that the number of identifiable paths is upper bounded by $K < 0.8187 M_t$.

In Algorithm 1, we show the detailed procedures for estimating multipath parameters using IMDF.

\begin{algorithm}[h]
	\caption{IMDF for DD-DP Parameter Estimation}\label{algorithm1}
	\begin{algorithmic}[1]
		\State Compute the least squares (LS) estimate of $\H$, i.e., $\hat\H_\text{LS}$, and form $\hat{\tH}^{(\rm p,q)}$ via \eqref{tildetH}.
		\State Use Theorem 2 to pre-calculate $\{P_x,P_y,P_r,Q_x,Q_y,$ $Q_r\}$, such that each $\tilde{\tH}^{(\rm p,q)}$ can be reshaped into a $P_xP_yP_r\times Q_xQ_yQ_r$ matrix which is denoted as $\tilde{\H}^{(\rm p,q)}$.
		\State Perform forward-backward smoothing on the conjugate of $\tilde{\H}^{(\rm p,q)}$ to obtain $\grave{\H}^{(\rm p,q)}$, and then $\forall \mathrm{p}\in\{\mathrm{(V_r,H_r)}\}$ and $\mathrm{q}\in\{\mathrm{(V_t,H_t)}\}$, stack $\{\tilde{\H}^{(\rm p,q)},\grave{\H}^{(\rm p,q)}\}$ into a $P_xP_yP_r\times 8Q_xQ_yQ_r$ matrix, denoted by $\check\H$.
		\State Perform 3-D IMDF to $\breve\H$ and obtain the estimates of $\{\theta_k, \varphi_k, \vartheta_k\}$.
		\State Use $\{\hat\theta_k, \hat\varphi_k, \hat\vartheta_k\}$ to construct $\hat\V_r,\hat{\V}_x,\hat{\V}_y$, and then estimate the path-loss matrix $\B$ via \eqref{eq:recal_B}
	\end{algorithmic}\label{Alg_IRLS}
\end{algorithm}


\section{Numerical Results}

Consider a MIMO system with an $4 \times 8$ DP URA at the BS and a $2$-element DP ULA at the MS. This particular case is of considerable practical interest in 3GPP as a candidate for implementation \cite{3gpp}. 
In the simulation, we assume that the multipath propagation gains are Rician distributed, and all the multipath parameters are randomly (uniformly) drawn. 
In the simulation, we assume that the multipath propagation gains are Rician distributed, and all the multipath parameters are randomly (uniformly) drawn. The BS covers $[0^\circ,90^\circ]$ elevation angular range and $(-45^\circ,45^\circ)$ azimuth angular range, while the MS only covers $[-60^\circ,60^\circ]$ azimuth angular range since the elevation angle is zero for ULA, i.e., $\theta_k\sim \mathcal{U}(-\pi/3,\pi/3),\,
\varphi_k\sim\mathcal{U}(0,\pi/2),\,  \vartheta_k\sim\mathcal{U}(-\pi/3,\pi/3)$. 
The non-parametric linear LS channel estimate is also plotted as a performance benchmark. 
All the results are averaged over 500 Monte-Carlo trials using a computer with 3.2 GHz Intel Core i5-4460 and 4 GB RAM.
The normalized MSE (NMSE) of channel estimates is computed from $\frac{1}{500}\sum_{i=1}^{500}\|\hat{\H}_{i}-\H\|_F^2/\|\H\|_F^2$
where $\hat{\H}_{i}$ denotes the estimate from the $i$th Monte-Carlo trial.

The number of multipath randomly varies from 1 to 6.  
Since the channel exhibits sparse property, we include a compressive sensing (CS) based technique \cite{jarvis} for comparison, 
where each angle is quantized with 7 bits, so the resulting dictionary is with size $4M_rM_t\times 2^{23}$, which however is infeasible in a conventional desktop. To make this algorithm work in a fast fashion, after obtaining the LS channel estimate, we reshape each sub-block of the channel estimate as an $M_r\times M_x\times M_y$ tensor and average them. Then we implement 3-D FFT with 128 points to estimate $\{\theta,\vartheta,\varphi\}$, following the so-called peak-picking technique. Finally, we update the path-loss matrix $\B$ via \eqref{eq:recal_B}.
We test the performance of all the competitors under known and unknown number of multipath. For the latter, we set $K=6$ to all the algorithms. Moreover, orthogonal pilots are employed.

\begin{figure}[t]
	\begin{center}
		\subfigure[known $K$]{\label{fig:mseknownChannel1} \includegraphics[scale=0.62, trim=0 0 0 0]{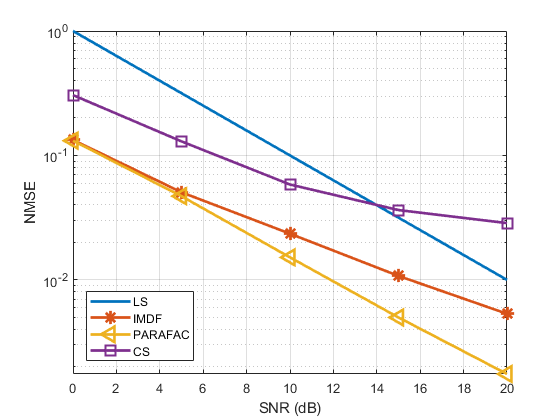}} 
		\subfigure[unknown $K$]{\label{fig:mseunkonwnChannel1} \includegraphics[scale=0.62, trim=0 0 0 0]{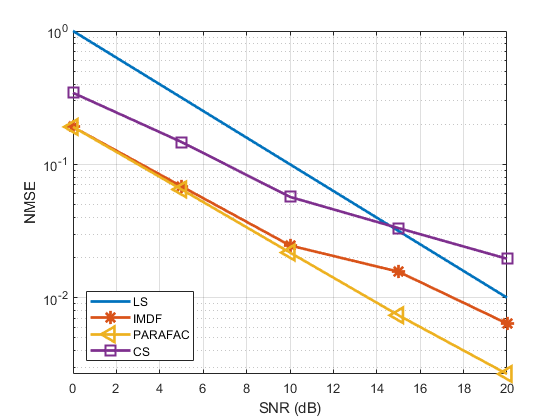}}
	\end{center}
\vspace{-1em}
	\caption{NMSE of versus SNR.}\label{msevssnr}
\end{figure}

It is observed from Fig. \ref{msevssnr} that PARAFAC outperforms the IMDF, LS and CS algorithms in both cases. Compared to Fig. \ref{fig:mseknownChannel1}, PARAFAC, IMDF and CS suffer slight performance loss in Fig. \ref{fig:mseunkonwnChannel1}, where the exact number of multipath is unknown. When SNR $>14$ dB, we see that the NMSE of CS is even worse than the LS method. This is mainly because as SNR increases, the performance of CS is limited by the resolution ability of dictionary.

\section{Conclusion}
We considered the parameter estimation problem for the DD channel model with DP arrays -- which is a setup that is of particular interest to standard organizations and industry. We proposed a tensor-based method to handle this challenging problem, which guarantees identifiability of the parameters of interest under mild and practical conditions.
We also proposed a reduced-complexity algorithm that is based on 3D harmonic retrieval to handle the same problem, with slight parameter accuracy loss but much faster runtime performance. Numerical simulations support our analysis and show that the proposed procedures, esp. the IMDF-based one, are very effective and promising for actual implementation.


\clearpage

\begin{thebibliography}{1}
	
    \bibitem{3gpp_stand}
	A. Kammoun, H. Khanfir, Z. Altman, M. Debbah and M. Kamoun, ``Preliminary results on 3-D channel modeling: From theory to standardization,'' \textit{IEEE Journal on Selected Areas in Communications}, vol. 32, no. 6, pp. 1219-1229, June 2014.
	
	
	
	\bibitem{3gpp}
	Z. Bai, ``Evolved universal terrestrial radio access (E-UTRA); physical layer procedures,'' \textit{3GPP, Sophia Antipolis, Technical Specification}, 36.213 v. 11.4.0, 2013.
	

	\bibitem{polarizationHeath}
	D. Zhu, J. Choi and R. W. Heath, ``Two-dimensional AoD and AoA acquisition for wideband mmWave systems with dual-polarized MIMO,'' \textit{IEEE Trans. Wireless Comm.}, , vol. 16, no. 12, pp. 7890-7905, Dec. 2017.
	
	\bibitem{polarization1}
	G. J. Foschini and M. J. Gans, ``On limits of wireless communications in a fading environment when using multiple antennas,'' \textit{Wireless Personal Communications}, vol. 6, no. 3, pp. 311-335, 1998.
	
	
	
	
	
	

	
	
	
	\bibitem{codebook4}
	A. Alkhateeb, G. Leus and R. W. Heath,``Limited feedback hybrid precoding for multi-user millimeter wave systems,'' \textit{IEEE Trans. Wireless Comm.}, vol. 14, no. 11, pp. 6481-6494, 2015.
	
	\bibitem{panos}
	P. N. Alevizos, X. Fu, N. Sidiropoulos, Y. Yang and A. Bletsas, ``Non-uniform directional dictionary-based limited feedback for massive MIMO systems,'' \textit{Proc. of 15th Inter. Symp. Modeling and Optimization in Mobile, Ad Hoc, and Wireless Networks (WiOpt),} Paris, pp. 1-8, 2017.
	
	\bibitem{jarvis}
	W.U. Bajwa, J. Haupt, A.M. Sayeed, and R. Nowak, ``Compressed channel sensing: A new approach to estimating sparse multipath channels,'' \textit{Proc. of the IEEE}, vol. 98, no. 6, pp. 1058-1076, 2010. 
	
	\bibitem{ce1}
	H. Lin, F. Gao, S. Jin and G. Y. Li, ``A new view of multi-user hybrid massive MIMO: Non-orthogonal angle division multiple access,'' IEEE J. Selected Areas in Comm., vol. 35, no. 10, pp. 2268-2280, Oct. 2017.
	
	
	
	\bibitem{polarized1}
	J. Li and R. T. Compton, ``Two-dimensional angle and polarization estimation using the ESPRIT algorithm,'' \textit{IEEE Trans. Antennas Propagat.}, vol. 40, pp. 550-555, 1992.
	
	\bibitem{Sid2017}
	N.D. Sidiropoulos, L. De Lathauwer, X. Fu, K. Huang, E.E. Papalexakis, and C. Faloutsos, ``Tensor decomposition for signal processing and machine learning'', \textit{IEEE Trans. Signal Process.}, vol. 65, no. 13, pp. 3551-3582, 2017
	
	\bibitem{liu}
	J. Liu and X. Liu, ``An eigenvector-based approach for multidimensional frequency estimation with improved identifiability,'' \textit{IEEE Trans. Signal Process.}, vol. 54, no. 12, pp. 4543-4556, 2006.

	
	
	\bibitem{parafac1}
	L. D. Lathauwer, B. D. Moor, and J. Vandewalle, ``Computation of the canonical decomposition by means of a simultaneous generalized schur decomposition,'' \textit{SIAM J. Matrix Anal. Appl.}, vol. 26, no. 2, pp. 295-327, 2004.
	
	
	\bibitem{parafac2}
	K. Huang, N. D. Sidiropoulos, and A. P. Liavas, ``A flexible and efficient algorithmic framework for constrained matrix and tensor factorization,'' \textit{IEEE Trans. Signal Process.}, vol. 64,no. 19, pp. 5052-5065, 2016.
	
	\bibitem{freqest1}
	C. Qian, L. Huang, H. C. So, N. D. Sidiropoulos, and J. Xie, `Unitary PUMA algorithm for estimating the frequency of a complex sinusoid,'' \textit{IEEE Trans. Signal Process.}, vol. 63, no. 20, pp. 5358-5368, 2015.
	
	\bibitem{freqest2}
	D. C. Rife and R. R. Boorstyn, ``Single tone parameter estimation from discrete-time observations,'' \textit{IEEE Trans. Inf. Theory}, vol. IT-20, no. 5, pp. 591-598, 1974.
	
	\bibitem{nikos1}
	N. Sidiropoulos and R. Bro, ``On the uniqueness of multilinear decomposition of $N$-way arrays,'' \textit{J. Chemometrics},
	vol. 14, no. 3, pp. 229-239, 2000.

	\bibitem{krank}
	N. D. Sidiropoulos, and X. Liu, ``Identifiability results for blind beamforming in Incoherent multipath with small delay spread,'' \textit{IEEE Trans. on Signal Process.}, vol. 49, no. 1, pp. 228-236, 2001.	
	
	
%
%
\end{thebibliography}
\end{document}